%% file: main.tex
\documentclass{article}
\usepackage{spconf,amsmath,graphicx}

\usepackage{multirow}
\usepackage{hhline}
\usepackage[caption=false]{subfig}

\usepackage[utf8]{inputenc} % allow utf-8 input
\usepackage[T1]{fontenc}    % use 8-bit T1 fonts
\usepackage{graphicx}
\usepackage{url}

\title{Mask Mining for Improved Liver Lesion Segmentation}

\name{Karsten Roth$^{1,2}$, Jürgen Hesser$^{1,2}$, Tomasz Konopczyński$^{1,2,3}$}
\address{$^1$Department of Data Analysis and Modeling in Medicine, Medical Faculty Mannheim\\
	     $^2$IWR, ZITI, Heidelberg University \hspace{10pt} $^3$Tooploox}

\begin{document}
%\ninept
%
\maketitle
\begin{abstract}
We propose a novel procedure to improve liver and lesion segmentation from CT scans for U-Net based models.
Our method extends standard segmentation pipelines to focus on higher target recall or reduction of noisy false-positive predictions, boosting overall segmentation performance.
To achieve this, we include segmentation errors into a new learning process appended to the main training setup, allowing the model to find features which explain away previous errors.
We evaluate this on semantically distinct architectures: cascaded two- and three-dimensional as well as combined learning setups for multitask segmentation.
Liver and lesion segmentation data are provided by the Liver Tumor Segmentation challenge (LiTS), with an increase in dice score of up to 2 points.
\end{abstract}
\begin{keywords}
U-Net, Liver Lesion Segmentation, Medical Imaging, Data Mining
\end{keywords}
\section{Introduction}
Liver imaging nowadays is mostly done via Computed Tomography (CT)~\cite{lits}. 
Providing fully-automatic segmentation of liver and lesion tissue from CT data can hence be a useful tool to help with diagnosis and treatment planning.
% Network Introduction
Common approaches utilize U-Nets~\cite{unet}, e.g.~\cite{nnunet,lits,multidice}.
% Problem Introduction
% However, as parameter-heavy neural network training has no closed-form solution, iterative procedures are used which make it hard to predict the network behaviour prior to training
However, training of neural networks can be a difficult endeavour.
To improve on existing scores, computationally expensive re-runs without guarantee of improvement are often needed.\\ 
% Even then, performance is never perfect, with new segmentation errors generated in each run.
% Proposal Introduction
We thus suggest a novel pipeline to reliably boost network segmentation performances by including segmentation errors as novel training masks in a post-training step.
\input{Figures/figure3.tex}
% Reference to related work
Prior work on the inclusion of segmentation errors into the training process include~\cite{tversky,tversky2}, who propose a Tversky-coefficient-based loss, which generalizes the standard Dice coefficient loss to include additional hyperparameters for penalizing false-positive or false-negative predictions during training.~\cite{fpgan} utilize segmentation error types in a complex adversarial setup, where refinement networks are trained on top of the basic setup to remove these errors. 
While the former introduce new hyperparameters, the usage of adversarial networks in \cite{fpgan} limits the usable network architectures. 
In both cases, heavy tuning and reruns are required for different architectural setups, as these methods are linked directly to the main learning process. This holds especially true going to three-dimensional data, which is common for many medical segmentation tasks.\\
We therefore propose to use segmentation error types in a setup separate to the main training.
Using segmentation errors of the learned networks, we append a secondary training process with specific loss functions to provide a framework that helps networks explain away own segmentation errors, thereby boosting segmentation performance (see fig.\ref{fig:expic} for qualitative impressions). Previous work s.a. \cite{mic} has shown the benefit of explaining away undesired properties. 
This means that our method stays independent of architecture and data choices, and allows for improved performance without costly reruns of the full setup.\\
Distinctly different networks and datatypes are tested to check the architecture- and datatype-independent applicability. This includes 2D and 3D data utilised in different training styles which are based around 2D and 3D U-Net~\cite{unet,unet3d,combined} pipelines (fig.~\ref{fig:allarchs}). Both training and evaluation is done on the Liver Tumor Segmentation (LiTS) dataset~\cite{lits}, showing consistent improvements in all setups.

\input{Figures/figure2.tex}
\input{Figures/figure1.tex}

\section{METHOD}
\label{sec:format}
Fundamental for our proposed extension (fig.~\ref{fig:fullpipe}) is the generation of new training masks to alter the current network performance and allow the network to learn from its own errors.
% \paragraph{Basic Setup}
\subsection{Basic Setup}
A segmentation pipeline of choice is trained until convergence following any training procedure.
Now, segmentation masks over the training data are generated through single forward passes with minimal computational burden.
These are then compared to the original ground truth to determine new training classes for each pixel, based on segmentation error cases:  
\textit{True Negative}, \textit{False Positive}, \textit{False Negative} and \textit{True Positive}. This gives four target classes compared to the binary case with two classes. 
We then append four single-layer output channels serving as error prediction layers to the output layer, introducing no relevant new parameters, but ensuring that all previously learned weights are kept until retraining on the novel masks is performed.
Due to the initial pretraining, convergence occurs much faster.
%Due to the initial pretraining, \textit{convergence occurs much faster}.
% \paragraph{Relevance of loss function}
\subsection{Relevance of loss function}
Assuming the majority of predicted pixels to be true positive or negative after training, we distinguish two approaches based on the choice of loss:\\
\emph{A pixel-weighted crossentropy loss (pwce)} (e.g.~\cite{unet}) gives highest learning signal to high frequency targets.
As we have a high imbalance towards true positive/negative predictions, retraining on error masks primarily reinforces these predictions while dropping noisy false positives.
The retrained multiclass error case predictions are then grouped into true positive/false negative and true negative/false positive predictions to generate a final binary segmentation mask:
\begin{equation}
    O^k_{ijm}(x^k) = \lfloor\frac{\text{argmax}_{c\in[0,..,C-1]}\text{ }{\phi^{multi}_{ijm,c}(x^k)}}{2}\rfloor\label{eq:conversion}
\end{equation}
%%%
\emph{A dice-coefficient based loss} (e.g. \cite{multidice}) injects a stronger learning signal for underrepresented classes for higher recovery of false-negative/positive pixels. 
Here, the primary interest lies in explaining away obfuscating features while retaining crucial ones, so the true positive error mask class is replaced with the ground truth segmentation mask. 
This allows the network to transfer properties generating false-positives to the respective output channel and recover generators for false-negative predictions. The final segmentation is taken directly from the true positive output channel.

Both loss functions offer a potential boost in performance and are mentioned for completeness.
%\textit{However, for all subsequent results, we utilize a dice-based loss as it provides marginally higher improvements.}
However, for all subsequent results, we utilize a dice-based loss as it provides marginally higher improvements.

\section{Application to Liver and Lesion Segmentation}
\label{sec:pagestyle}

\subsection{Network Architectures}
%We investigate the performance of our method on liver and lesion segmentation by evaluating dice score performance on distinct architectures: \textit{(i)} \textit{Cascaded 2D}\cite{cascaded}, which trains a 2D segmentation network for liver and lesion segmentation separately, \textit{(ii)} \textit{Cascaded 3D}, which does the same for a 3D setup and \textit{(iii)}, \textit{Combined Cascaded 2D}\cite{combined}, which trains separate segmentators for liver and lesion in a simultaneous setup. All networks are equipped with common extensions such as multislice inputs\cite{han}, batch normalization\cite{batchnorm}, residual blocks\cite{residual} and squeeze-and-excitation modules\cite{squeeze}. 
We investigate the performance of our method on liver and lesion segmentation by evaluating dice score performance on distinct architectures: Cascaded 2D~\cite{cascaded}, which trains a 2D segmentation network for liver and lesion segmentation separately, Cascaded 3D, which does the same for a 3D setup and Combined Cascaded 2D~\cite{combined}, which trains separate segmentators for liver and lesion in a simultaneous setup. All networks are equipped with common extensions such as multislice inputs~\cite{han}, batch normalization~\cite{batchnorm}, residual blocks~\cite{residual} and squeeze-and-excitation modules~\cite{squeeze}. 
Each pipeline is trained to convergence before applying our extension to ensure that we do not just prolong the training process. 
For liver segmentation, initial training is done with pwce loss $L^{pwce}$ and distance-transformation weightmaps (see~\cite{unet}).The lesion segmentation loss is based on dividing $L^{pwce}$ by a smooth dice score $L^{dice}$ (see e.g. \cite{multidice}), 
$L_{combined} = L^{pwce} \cdot \left(L^{dice} +\epsilon \right)^{-1}\label{eq:lesion}$, with $\epsilon=10^{-5}$.
% , for lesion segmentation:
% with images $\{x^k\}_{k\in[1,K]}$ in minibatch of size $K$, network $\phi$, target mask $t^k$ and weightmap $w^k$ with width $W$ and height $H$. For numerical stability, a small $\epsilon=10^{-5}$ is added. 
% Schematics are visualized in fig. \ref{fig:allarchs} for all three pipelines.
% \paragraph{LiTS dataset}
\subsection{LiTS dataset}
The Liver Tumor Segmentation (LiTS) dataset~\cite{lits} contains 131 3D lower abdominal CT scans with liver/lesion ground truth masks, as well as 70 test volumes evaluated by online submission. The dataset is publicly (Creative Commons License) and was collected for ISBI/MICCAI 2017.
All volumes have horizontal dimensions of 512 with near constant resolution. In the axial direction dimensionality and resolution vary strongly, which is a relevant factor for any approach using higher-than-two dimensional data input.
Before training, the data is bounded to $[-100,600]$ $HU$ before performing normalization.
For evaluation, only the largest connected component is used to generate the final liver segmentation.
% \paragraph{Results}

\subsection{Implementation Details}
The full pipeline is implemented with PyTorch~\cite{pytorch}. We use a $85\%|15\%$ train/val split and run everything on a NVIDIA GeForce 1080Ti.
$256\times256$ crops with batchsize $12$ are used for 2D training and $128\times128\times64$ crops with batchsize $2$ 3D training. For liver segmentation, crops are taken randomly, while for lesion segmentation crops in and around the liver are used. Standard data augmentation using random horizontal and vertical flips, random rotation and random zooming is performed, all in axial direction. For optimization, Adam~\cite{adam} with an initial learning rate of $10^{-5}$ and $L2$-regularisation $\lambda=10^{-5}$ is used. Standard step-based learning rate scheduling is included as well. Training is performed for 70 epochs to ensure convergence, saving the best validation weights.

\subsection{Results}
We compute the averaged dice score per test volume before and after application of our method for all architectures. Here, relative improvement is the key metric to examine.
Results are summarized in tab.~\ref{tab:performance}, showing a consistent gain over the initially trained model, especially for the combined training setup. This is arguably due to the simultaneous boost in liver and lesion segmentation performance.
The inclusions of mined trained masks into the training process specifically benefits validation performance. This is rooted in the splitting procedure, as training and validation set are drawn from the same sample set. Due to different sources contributing to the dataset~\cite{lits}, the test set samples therefore differ much stronger from the training set. Newly mined features are hence more expressive on the validation set.

\input{Figures/figure4.tex}
\input{Tables/performance.tex}

\section{CONTROL OF SEGMENTATION ERRORS}
We also qualitatively study the usage of our method to control the produced segmentation error types, with examples in fig.~\ref{fig:expic}. To do so, the distribution of segmentation error types before and after running a mask mining step with either a multiclass dice loss or a multiclass pwce loss is compared over all proposed architectures, see fig.~\ref{fig:control}. A clear shift in false-positive and false-negative pixels depending on the choice of utilized loss can be seen compared to the initial pre-mask-mining setup. The network segmentation behaviour changes for different loss functions, while the segmentation performance in both cases is improved.

\section{Conclusion}
We introduced a novel extension to standard liver and lesion segmentation pipelines on the basis of the Liver Tumor Segmentation (LiTS) dataset.
By helping the network learn and thereby explain away previously made errors using automatically generated training labels, we boost segmentation performance on different and distinct architectures and training styles. 
% Konops's
% What is the potential of the method
Although we present our work on the task of liver and liver lesion segmentation from CT scans via deep learning U-net like architectures,
due to the architecture-independent applicability our method can be extend to other medical image segmentation problems.
Not only to a variety of other applications but also to other machine learning based semantic segmentation techniques.
We see our Mask Mining idea as an addition to boosting and ensemble methods.
% What is the achieved gain versus the SOTA of the method
Decreasing the number of false negatives is of great significance, especially in medical image analysis. Most significantly, the use of our Mask Mining method allows potential detection of previously omitted objects of interest.
% What is the limitation of the method
However, we still observe limitation in terms of the dice score. A straightforward idea for improvement would be an iterative approach in which the Mask Mining could be repetitively used on the learning model.
It should help the model to further increase the sensitiveness to the errors performed by the model.
We leave this issue for future research and investigation.

\bibliographystyle{IEEEbib}
\bibliography{egbib}

\end{document}

%% file: Figures/figure3.tex
\begin{figure}[ht]
\begin{center}
   \includegraphics[width=1\linewidth]{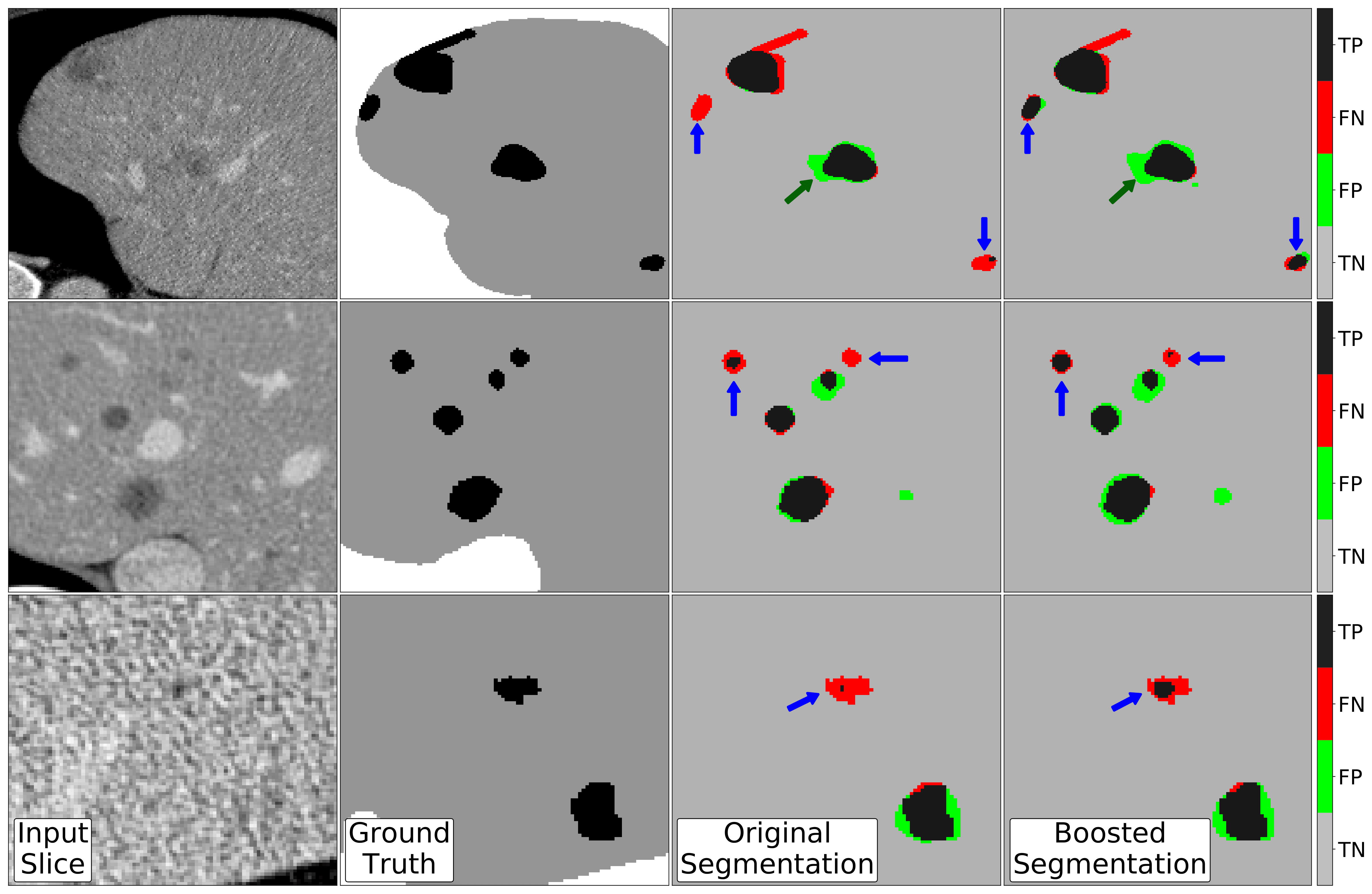}
\end{center}
\vspace{-10pt}
   \caption{
   %\textit{Qualitative evaluation before and after mask mining.}
   Qualitative evaluation before and after mask mining.
   \textit{Original Segmentation} denotes initial,  and \textit{Boosted Segmentation} the performance after retraining, with clearly reduced false negatives (\textit{red}) and few new false positives (\textit{green}). \textit{Grey} and \textit{black} denote true negative/positive errors.}
\label{fig:expic}
\vspace{-10pt}
\end{figure}

%% file: Figures/figure2.tex
\begin{figure*}[htb]
   \begin{center}
   \includegraphics[width=1\linewidth]{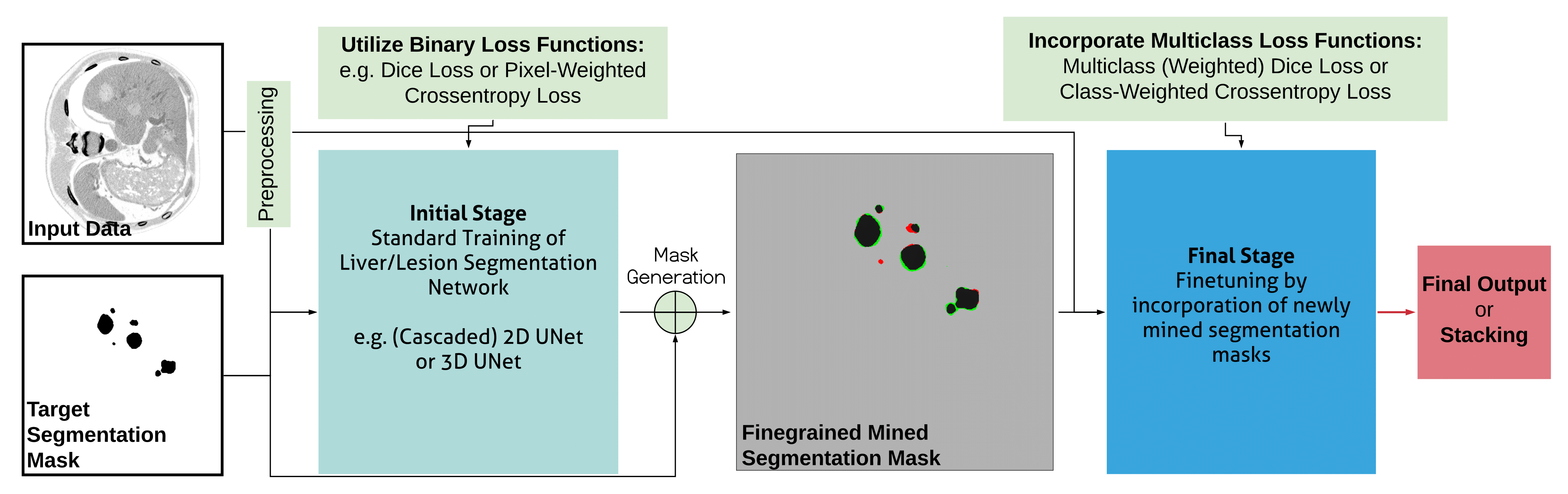}
   \end{center}
   \caption{
   %\textit{The Mask Mining Pipeline.}
   The Mask Mining Pipeline.
   Starting from the original training setup, generated segmentation masks are compared with the ground truth masks to generate new finegrained multiclass training masks containing previously made segmentation errors. This allows the network to learn to explain away mistakes.}
\label{fig:fullpipe}
\end{figure*}

%% file: Figures/figure1.tex
\begin{figure*}[htb]
    \begin{center}
  \includegraphics[width=1\linewidth]{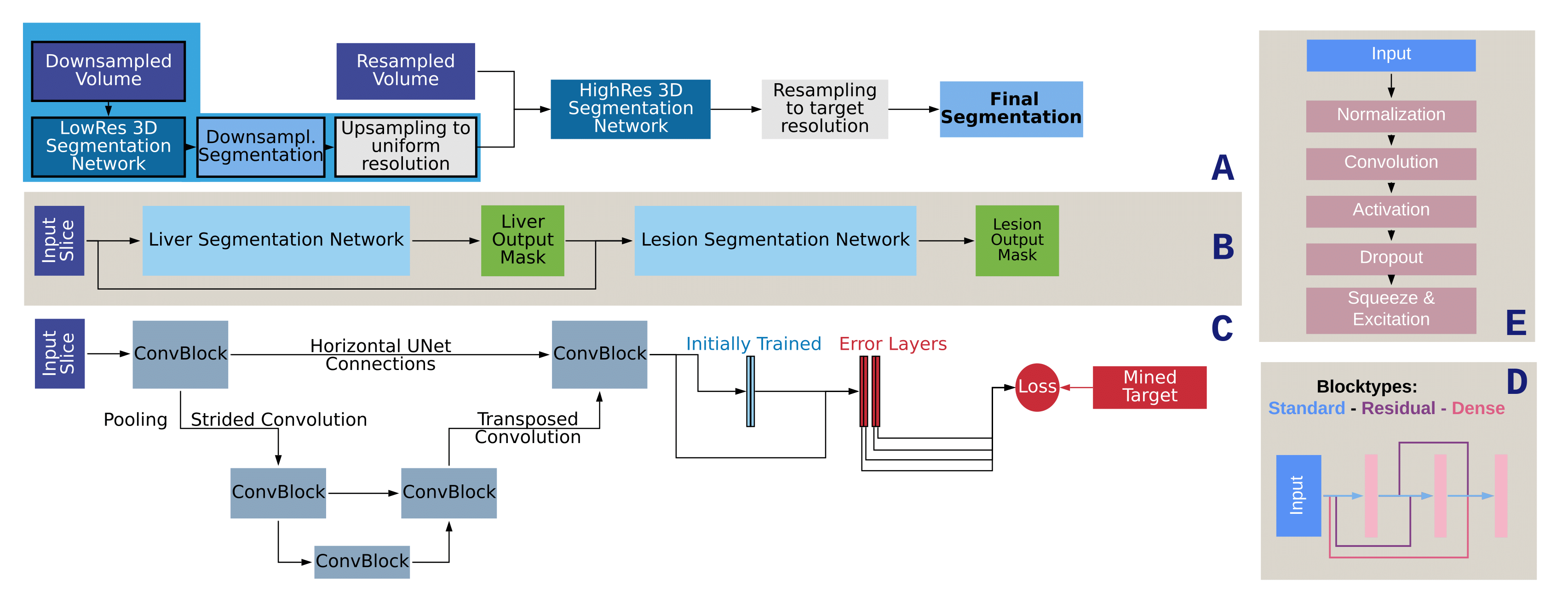}
    \end{center}
  \caption{
  %\textit{Schematics of all utilised architectures and training pipelines.} 
  Schematics of all utilised architectures and training pipelines.
  \textbf{(A)} (Optional two-step) pipeline for 3D liver/lesion segmentation networks. \textbf{(B)} Setup for simultaneous training of liver and lesion segmentation networks. \textbf{(C)} Basic U-Net architecture, including the additional kernelsize 1 error layers to train on error masks. \textbf{(D)} Convolutional U-Net Block with either dense, residual or basic connectivity. \textbf{(E)} Layers utilized to represent a single  ConvBlock-Layer (pink in (D)).}
\label{fig:allarchs}
\end{figure*}

%% file: Figures/figure4.tex
\begin{figure*}[ht]
\begin{center}
   \includegraphics[width=1\linewidth]{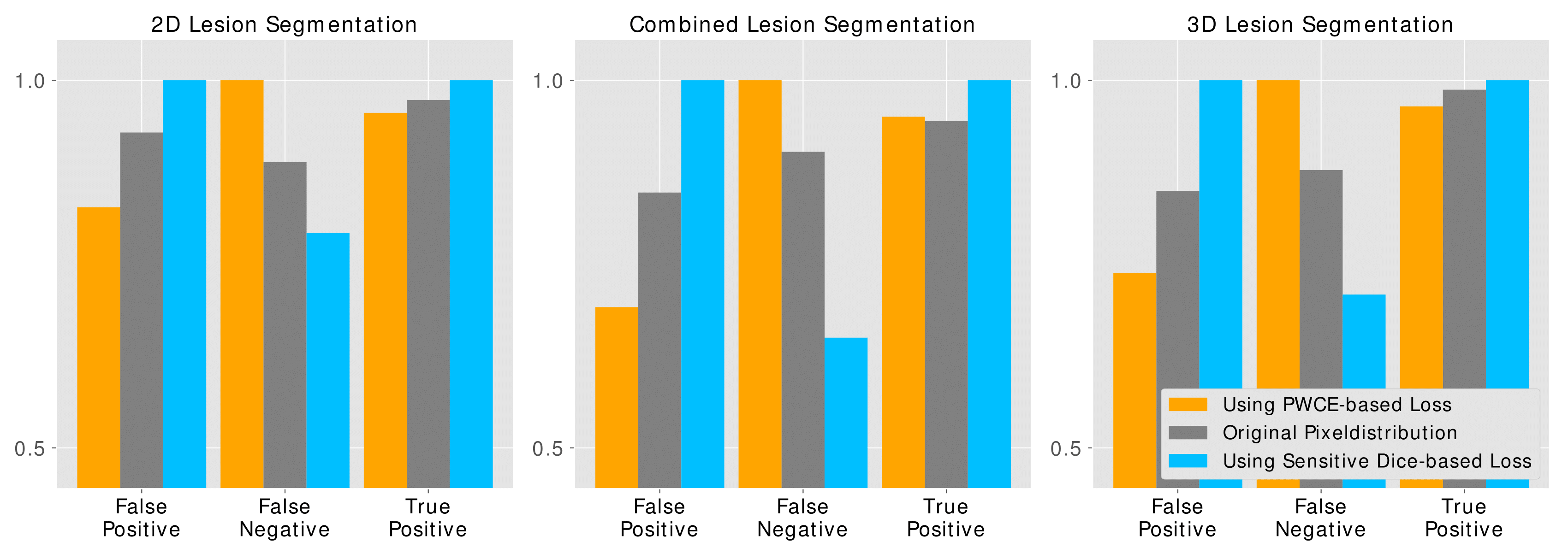}
\end{center}
   \caption{
   %\textit{Qualitative examination of the control capability of our pipeline extension.}
   Qualitative examination of the control capability of our pipeline extension.
   We evaluate false positive/false negative/true positive pixel count change for a fixed validation set on lesion segmentation capability for all architectures. Each bar group is normalized to the highest group value for visual clarity. We see that the network performance can be tweaked retrospectively for higher recall (blue) or robustness (orange) without adding new hyperparameters and independent of the underlying architecture.}
\label{fig:control}
\end{figure*}

%% file: Tables/performance.tex
\begin{table*}[t!]
\caption{
%\textit{Quantitative evaluation of network performance before and after error inclusion (Inc.)}.
Quantitative evaluation of network performance before and after error inclusion (\textit{Inc.}).
We show volume-averaged dice scores for liver and lesion segmentation on the test set and fixed training and validation sets. We see a clear improvements in dice scores. In addition, error inclusion reduces seed-dependent variation in performance (measured over three runs).} 
\centering
\begin{tabular}{|l|c|c|c|c|c|c|}
\hline
\multirow{2}{*}{\textbf{Setup }} & \multicolumn{2}{c|}{Training Dice} & \multicolumn{2}{c|}{Validation Dice} & \multicolumn{2}{c|}{Online Test Dice}\\
\cline{2-7}
& Liver & Lesion & Liver & Lesion & Liver & Lesion\\
 
\hhline{=======}
\textbf{2D} & $96.9 \pm 0.3$ & $71.9 \pm 0.4$ & $95.9 \pm 0.3$ & $63.5 \pm 0.6$ &  $95.3 \pm 0.2$ & $62.9 \pm 0.3$\\  
\hline
\textit{Inc.} & $\mathbf{97.0 \pm 0.1}$ & $\mathbf{73.7 \pm 0.2}$ & $\mathbf{96.3 \pm 0.2}$ & $\mathbf{64.9 \pm 0.2}$ & $\mathbf{95.5 \pm 0.3}$ & $\mathbf{63.5 \pm 0.2}$\\ 
\hhline{=======}
\textbf{3D} & $92.2 \pm 1.4$ & $63.0 \pm 0.8$ & $91.4 \pm 0.9$ & $56.8 \pm 2.0$ & $91.2 \pm 1.0$ & $55.5 \pm 0.9$\\ 
\hline
\textit{Inc.} & $\mathbf{94.2 \pm 0.3}$ & $\mathbf{66.1 \pm 0.4}$ & $\mathbf{91.8 \pm 0.6}$ & $\mathbf{57.7 \pm 0.4}$ & $\mathbf{92.0 \pm 0.4}$ & $\mathbf{56.5 \pm 0.2}$\\ 
\hhline{=======}
\textbf{Cmb} & $94.5 \pm 0.3$ & $70.1 \pm 0.5$ & $92.9 \pm 0.7$ & $61.6 \pm 0.5$ & $93.4 \pm 0.3$ & $61.9 \pm 0.2$\\
\hline
\textit{Inc.} & $\mathbf{96.2 \pm 0.5}$ & $\mathbf{72.3 \pm 0.4}$ & $\mathbf{94.0 \pm 0.3}$ & $\mathbf{63.4 \pm 0.4}$ & $\mathbf{94.7 \pm 0.3}$ & $\mathbf{63.0 \pm 0.1}$\\
\hline

\end{tabular}
\label{tab:performance}
\end{table*}